\documentclass[aps,prl,reprint,superscriptaddress,super,superbib,amsmath,showkeys,notitlepage]{revtex4-1}
\usepackage{color}
\usepackage[pdftex]{graphicx}  
\newcommand{\ndg}{{\phantom{\dagger}}}
\newcommand{\dg}{\dagger}

\begin{document}

\title{Wigner time delay induced by a single quantum dot}

\date{\today}
\author{Max Strau\ss}
\affiliation{Insitut f\"ur Festk\"orperphysik, Technische Universit\"at Berlin, D-10263 Berlin, Germany}
\author{Alexander Carmele}
\affiliation{Institut f\"ur Theoretische Physik, Technische Universit\"at Berlin, D-10263 Berlin, Germany}
\author{Julian Schleibner}
\affiliation{Institut f\"ur Theoretische Physik, Technische Universit\"at 
Berlin, D-10263 Berlin, Germany}
\author{Marcel Hohn}
\affiliation{Insitut f\"ur Festk\"orperphysik, Technische Universit\"at Berlin, 
D-10263 Berlin, Germany}
\author{Christian Schneider}
\affiliation{Technische Physik, Physikalisches Institut,Wilhelm Conrad R\"ontgen Center for Complex Material  Systems, Universit\"at W\"urzburg, D-97074 W\"urzburg, Germany}
\author{Sven H\"ofling}
\affiliation{Technische Physik, Physikalisches Institut,Wilhelm Conrad R\"ontgen Center for Complex Material  Systems, Universit\"at W\"urzburg, D-97074 W\"urzburg, Germany}
\affiliation{SUPA, School of Physics and Astronomy, University of St. Andrews, St. Andrews KY16 9SS, United Kingdom}
\author{Janik Wolters} 
\affiliation{Insitut f\"ur Festk\"orperphysik, Technische Universit\"at Berlin, D-10263 Berlin, Germany}
\affiliation{Present address: Department of Physics, University of Basel, Klingelbergstrasse 82, CH-4056 Basel, Switzerland}
\author{Stephan Reitzenstein}
\affiliation{Insitut f\"ur Festk\"orperphysik, Technische Universit\"at Berlin, D-10263 Berlin, Germany}

\email[Correspondence and requests for materials should be addressed to S.R., e-mail address: ]{stephan.reitzenstein@physik.tu-berlin.de}

\begin{abstract}
Resonant scattering of weak coherent laser pulses on a single two-level system (TLS) realized in a semiconductor quantum dot is investigated with respect to a time delay between incoming and scattered light. This type of time delay was predicted by Wigner in 1955 for purely coherent scattering and was confirmed for an atomic system in 2013 [R. Bourgain et al., Opt. Lett. 38, 1963 (2013)]. In the presence of electron-phonon interaction we observe deviations from Wigner's theory related to incoherent and strongly non-Markovian scattering processes which are hard to quantify via a detuning-independent pure dephasing time. We observe detuning-dependent Wigner delays of up to 530\,ps in our experiments which are supported quantitatively by microscopic theory allowing for pure dephasing times of up to 950\,ps.
\end{abstract}

	
\maketitle

Scattering of light on the level of single photons and single emitters is heavily investigated to shed light on the underlying principles governing light-matter interaction on a microscopic and fundamental level. Of particular relevance are studies on resonantly driven two-level systems (TLSs) which allow one to explore predicted quantum optics effects in experiment and to refine their theoretical description. In this context it is very interesting to consider not only ''clean'' atomic systems but also solid state two-level emitters in the presence of decoherence. Here questions arise to which extend quantum optical effects can also be observed for TLSs interacting with the host matrix and how the coupling via phonons influences the underlying physics as well as and the atom-like character of the corresponding solid-state qubit.  

First experiments on resonantly excited solid state TLSs were performed in the limit of very low Rabi frequencies, also called the Heitler regime~\cite{Heitler1954}, where light scattering by an ideal TLS is dominated by elastically scattered photons, the resonant Rayleigh scattering. It has been shown that the radiation emitted by a semiconductor quantum dot (QD) in this regime shows distinct sub-Poissonian statistics~\cite{Muller2007, Ates2009}, while inheriting the first order coherence properties from the excitation source providing unparalleled coherence times~\cite{Matthiesen2012}. These exceptional properties have motivated further research on the coherent light scattering from QDs which includes the filtering of single photons from a coherent laser beam~\cite{Bennett2016, DeSantis2017}, a nanophotonic optical isolator~\cite{Sayrin2015}, and the generation of transform limited~\cite{Kuhlmann2015} and phase locked quantum light~\cite{Matthiesen2013a}. Moreover, coherent light 
scattering from solid state TLSs is relevant for applications photonic quantum technologies \cite{Proux2015,Bennett2015} and for interfacing hybrid quantum systems \cite{Meyer2015,Jahn2015,Wolters2017}. Experiments in the Heitler regime were complemented by studies on the Hong-Ou-Mandel effect at higher excitation strenght
to prove the efficient generation of single photons with a high degree of single-photon indistinguishability~\cite{Somaschi2016, Ding2016, Unsleber:16, Kreinberg2018}.  

Interestingly, in the Heitler regime, one can also observe a temporal delay between the exciting and the scattered pulse. Caused by the phase lag induced by the two-level system, a coherently scattered pulse is, in the absence of pure dephasing processes, maximally delayed by twice the radiative lifetime $T_1$. This effect is called the Wigner time delay and was first derived by Eugene Wigner in 1955 for scattering processes occurring in high energy physics \cite{Wigner1955a}. In the optical domain, this effect has been studied so far only for single atoms confined in a trap \cite{Bourgain2013}. In that case, a time delay up to 42 ns was reported which is fairly close to the theoretical limit of two times the spontaneous emission lifetime (26 ns) of the $^{87}$Rb atom's studied transition. 

In this Letter, we report on the scattering of weak coherent laser pulses on a single semiconductor QD. Interestingly, and in contrast to atomic systems, QDs are subjected to pure dephasing effects which lead a significant contribution of incoherently scattered photons and, as a consequence, to clear deviations from the expected behaviour for an ideal two-level emitter which are not explainable by simple Markovian decoherence theory. We observe detuning dependent delays of up to 530\,ps where the detuning dependence of the Wigner time delay is explained by considering non-Markovian electron-phonon 
effects within a microscopic theory. Therefore, the deviations from the ideal case, i.e. a Wigner limit of $2T_1$, are a very good measure to 
quantify and compactify the coherence properties of the QD without relying on a phenomenological derived pure dephasing time $T_2^*$. 

\begin{figure}[thbp]  
\centering
\includegraphics[width=\linewidth]{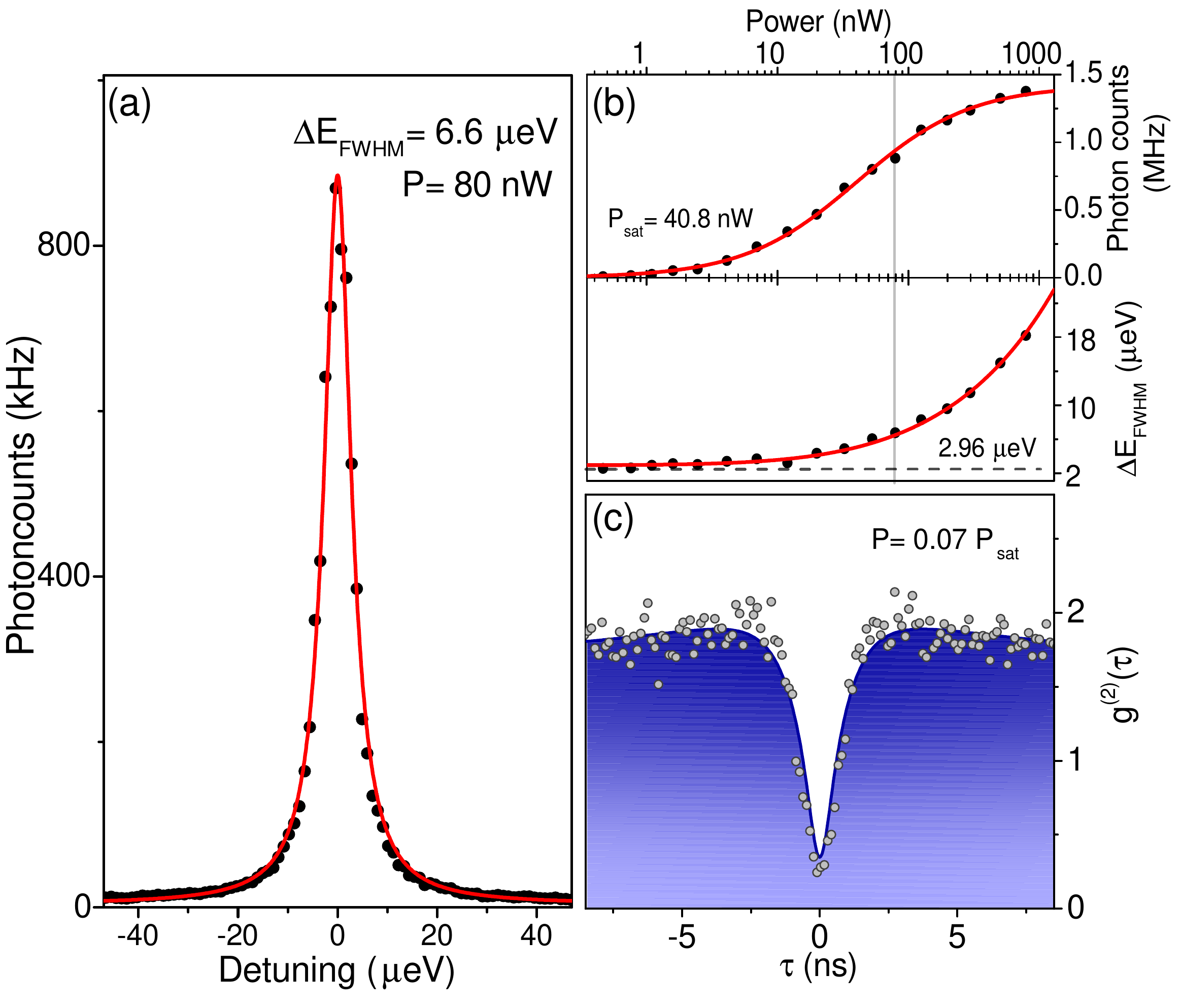}
\caption{Optical characterization of a single InGaAs QD under strict resonant excitation. (a) Fluorescence signal as a function of laser frequency at an excitation power of 80\,nW. The experimental data is well described by a Lorentzian lineshape with a width 6.6 $\mu$eV (FWHM) (red trace). (b) Upper (lower) panel: Power dependent measurements of fluorescence intensity (linewidth). The minimum observed linewidth is ($2.96\pm 0.11$)\,$\mu$eV. (c) Intensity auto-correlation measurement at 0.07\,$P_{\text{sat}}$.}
\label{graph:1}
\end{figure} 

The solid state two-level system studied in this work is formed by a self-assembled InGaAs QD. An ensemble of such QDs with a density of about  $10^9~\mathrm{cm}^{-2}$ is embedded into an epitaxially grown GaAs heterostructure consisting of a lower distributed Bragg reflector (DBR) with 24 $\lambda/4$-thick AlAs/GaAs mirror pairs, a central one-$\lambda$ thick GaAs cavity and an upper AlAs/GaAs DBR with 5 mirror pairs. Target QDs are self-aligned to photonic defects within the microcavity sample which enhances their photon extraction efficiency up to 42\% \cite{Maier2014} and thereby facilitates a comprehensive study of the Wigner time delay also at small excitation powers and large spectral detunings. For more details on the sample technology we refer to the supplementary information (SI).  

We resonantly excite our sample which is mounted inside a helium flow cryostat at 5\,K using a tunable CW diode laser. In addition, the sample is illuminated through the same beam path by a low intensity ($<100$\,pW) non-resonant diode laser ($\lambda=785$\,nm) to fill adjacent charge traps, thus narrowing the QD emission linewidth and stabilizing the fluorescence signal \cite{Nguyen2013a,gazzano2018}. Using a microscope objective (NA=0.65) the collimated, linearly polarized laser beam is focused on to the sample to a diffraction limited spot size of approximately 1~$\mu$m. A combination of a $\frac{\lambda}{4} $-plate, polarizing beam splitters and spatial filtering by coupling into a single mode fiber \cite{Kuhlmann2013} suppresses the collection of reflected laser light by the same objective. Typical light densities applied in this work are 1-10 W cm$^{-2}$ (10 mW cm$^{-2}$) of resonant (non-resonant) excitation. In order to observe the Wigner time delay, the cw-laser is amplitude modulated at a pulse repetition rate of 50 MHz using a fiber-based electro-optic modulator (EOM). Here, the exciting pulse is detected as time reference by tuning the laser off resonance and rotating the $\frac{\lambda}{4}$-plate by 0.1 degree with respect to the dark field configuration. Since the (non-modulated) non-resonant laser is two to three orders weaker than the resonant laser and operates in CW excitation we do not expect any influence of the non-resonant laser on the Wigner dynamics presented in this work. The luminescence is detected with an overall timing resolution of $\Delta \tau_{\text{FWHM}}=390$\,ps. More information about the experimental setup including a sketch is provided in the SI.   

Fig.~\ref{graph:1} shows results of the optical pre-characterization of the QD under study. The data were obtained for a charged exciton emitting at a wavelength of 924.9 nm (see SI for a micro-photoluminescence spectrum) from which we extract the parameters for simulating the Wigner time-delay experiment presented below. A laser scan across the exciton transition at an excitation power of 80\,nW is shown in Fig.~\ref{graph:1}~(a). The maximum intensity on resonance as a function of excitation power is plotted in the upper panel of Fig.~\ref{graph:1}~(b). Above saturation, we detect a maximum of $1.3*10^{6}$ photons/s on our single photon counting module (SPCM) on resonance. At an excitation power of 1 $\mu$W this corresponds to a fraction of extracted photons to incoming photons of $3 \times 10^{-7}$. The saturation behaviour of a TLS on resonance is given by $I \propto \rho_{11}=1/2\,S/(1+S)$. Here, $S$ denotes the saturation parameter defined on resonance as $S:=\Omega^2 T_1 T_2=P/P_{\text{sat}}$ where $P$ is the excitation power and $P_{\text{Sat}}$ the saturation power. From these measurements we also determine the dependence of the linewidth (FWHM) on the excitation power which is displayed in the lower panel of Fig.~\ref{graph:1}~(b). By fitting the expected behaviour of a TLS given by $\Delta \omega = 2/T_2 (1+S)^{1/2}$ to the data we extract a $T_2$ time of ($445 \pm 16$)\,ps. The radiative lifetime $T_1$ of the exciton is obtained from the resonant time resolved intensity auto-correlation measurement displayed in Fig.\,\ref{graph:1}\,(c).The extracted $T_1$ time is ($750\pm 150$)\,ps which corresponds to a natural linewidth of approximately 0.84\,$\mu \text{eV}$. We would like to note that in addition to the expected antibunching on timescales shorter than the lifetime of the emitter we observe a bunching on longer timescales which is most likely caused by blinking of the QD~\cite{Davanco2014}. We refer to the SI for resonance fluorescence spectra of the resonantly driven QD. 

After having obtained a good understanding of its emission properties, we now turn to the study of the Wigner time delay in our solid state TLS. In the limit of low Rabi frequencies and long pulses ( $\Delta t_{FWHM} \gg T_1$ and $P \ll P_{\text{sat}}$) the phase lag between scattered and exciting radiation leads to a time delay between the exciting and scattered pulse. This can be also understood as each Fourier component of the pulse is multiplied by a phase in the frequency domain which leads to a shift in the temporal domain. Since the phase lag is frequency dependent, a frequency mismatch is effectively turned into a temporal delay. To study the corresponding Wigner time delay we set the excitation power to a value of $P \sim 0.02\,P_{\text{sat}}$ under pulsed excitation (pulse repetition rate: 50 MHz, pulse width: 1.05 ns). 

\begin{figure}  
\includegraphics[width=\linewidth]{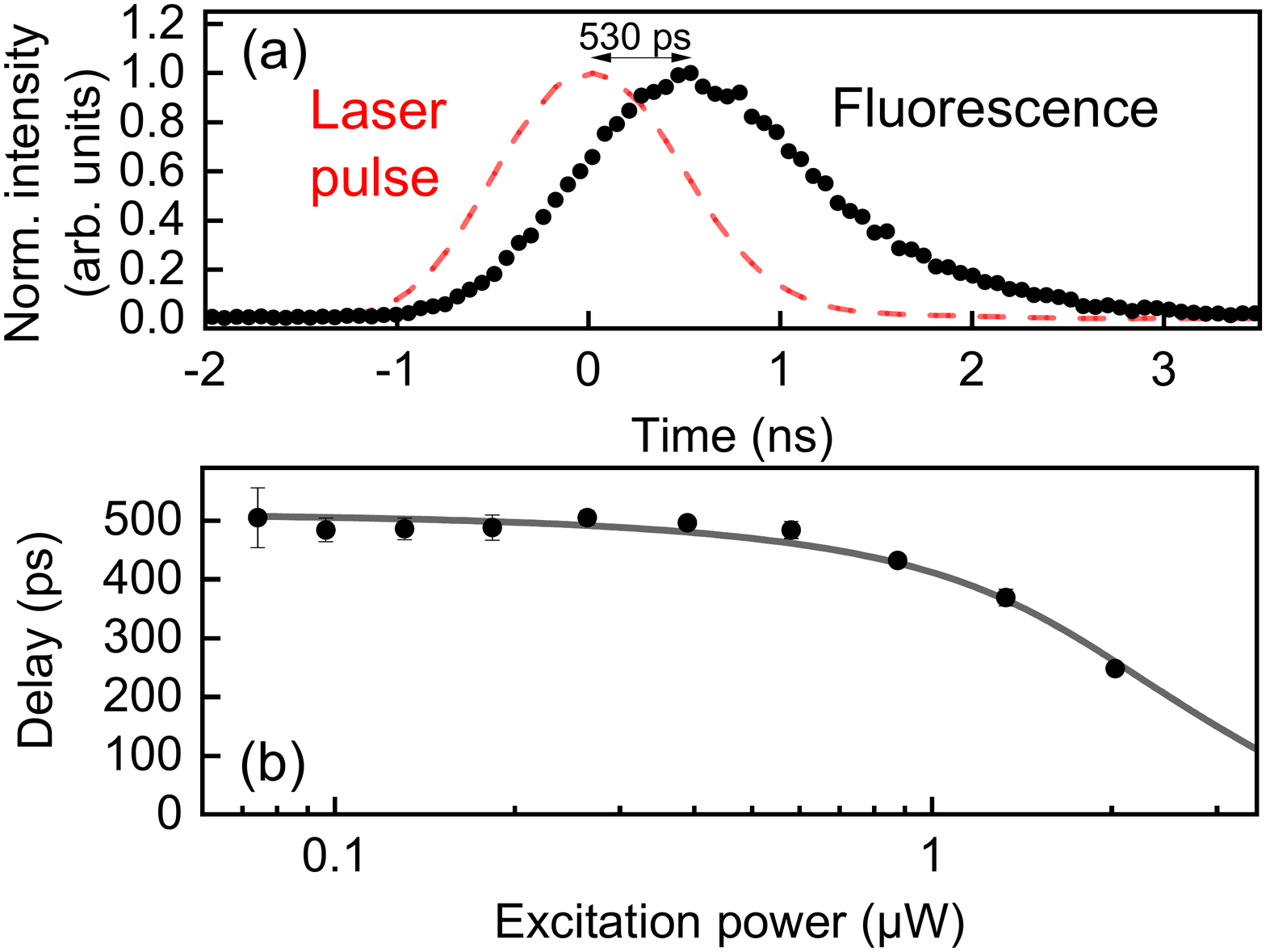}
\caption{(a) Red: Laser pulse used for exciting the QD. Black: Pulse scattered by QD for a laser detuning of $\Delta = $ 0.5\,$\mu$eV. The observed delay between the pulses is 530\,ps. (b) Time delay as a function of excitation power for zero detuning between the laser and the TLS. The experimental data (black bullets) shows a pronounced decrease of the time delay for excitation powers exceeding 600 nW which is well described by our microscopic theory (black line) taking non-Markovian dynamics into account.}
 \label{graph:4}
\end{figure}

Under these excitation conditions the generated pulses allow us to measure a significant Wigner delay as presented in Fig.~\ref{graph:4}~(a). Here, the exciting 
and scattered pulses are displayed for a small detuning of $\Delta =0.5\,\mu$eV. The center of the scattered pulse is shifted in time by a delay of 530\,ps - 
corresponding to the Wigner time delay of our solid state TLS - with respect to the exciting pulse. Importantly, the delay does increase if the intensity is lowered which indicates that we operate in the Heitler regime. This aspect, namely the excitation power dependence of the Wigner time delay, is studied in more 
detail on resonance in Fig.~\ref{graph:4}~(b). Here, the measured delay is plotted for an excitation power in the range of 74 nW to 2.0 $\mu$W. Clearly, we observe a 
delay of about 500 ps at low  excitation powers. More importantly, however, the Wigner delay saturates for low excitation power which means the Heitler regime is indeed reached. With increasing excitation power, the Wigner time delay stays at a plateau of approximately 500 ps up to about 600 nW above which it decreases significantly to a value of 250 ps at a power of 2.0 $\mu$W. The experimental result is confirmed and in very good quantitative agreement with our theory which considers non-Markovian processes and is introduced in the follwowing.

Our theory is based on the semiconductor Bloch equation \cite{kira2011semiconductor,carmele2010antibunching,richter2009two}. The population dynamics of the 
conduction band is derived in a density matrix approach $\left\langle a^\dagger_c a^{\phantom{\dagger}}_c\right\rangle=\text{Tr}\left[\rho a^\dagger_c a^{\phantom{\dagger}}_c \right]$ where
$\rho$ is the density matrix.
The dynamics of the density matrix is 
based on the Liouville-von Neumann equation $\partial_t \rho = -i[H/\hbar,\rho] +\Gamma \mathcal{D}[a^\dagger_v a^{\phantom{\dagger}}_c]\rho$ with the system Hamiltonian $H$ and the Lindblad superoperator $\mathcal{D}[J]\rho=2J\rho J^\dagger-\lbrace J^\dagger J, \rho \rbrace$ for the radiative decay, annihilating a conduction band and creating a valence band electron: $a^{\phantom{\dagger}}_c a^\dagger_v$ \cite{kira2011semiconductor}. 
The system Hamiltonian includes the
electron-phonon interaction and reads:
\begin{align}
H/\hbar =& 
\Delta a^\dg_c a^\ndg_c
+
\Omega(t) 
\left( a^\dg_v a^\ndg_c + a^\dg_c a^\ndg_v \right) \\ \notag
&+
\sum_{\bf q} 
\omega_q
b^\dg_{\bf q} b^\ndg_{\bf q}
+
\sum_q 
\left( g^{\bf q}_{vc} b^\ndg_{\bf q} + g^{{\bf q}*}_{vc} b^\dg_{\bf q} \right)
a^\dg_c a^\ndg_c
\end{align}
for a rotating frame in the laser frequency and detuning $ \Delta$ between laser and band gap transition frequency.
The coupling to acoustic phonons with
annihilation(creation) operator $b^{(\dg)}_{\bf q}$ and wave number ${\bf q}$ is incorporated via the coupling element $g^{\bf q}_{vc}$, and the linear dispersion reads: $ \omega_q = c_s |{\bf q}|$.
The electron-phonon interaction allows to include non-equilibrium scattering contribution but also leads to a hierarchy problem which is solved via the cluster-expansion approach \cite{richter2009two,carmele2013stabilization,daniels2011quantum} (please see the SI for the full set of equations of motion and material parameters).
The decrease of the Wigner delay time with increasing excitation in Fig.~\ref{graph:4}(b)
results from the fact that for larger excitation powers Rabi oscillations start to emerge. 
The stronger the TLS is driven, the earlier in time the maximum of the occupation density is reached, possibly even before the pulse maximum is reached. 
Therefore, the Wigner delay is only a good figure of merit for weak excitation, when the excitation power, or the pulse area does not determine the time lag and depends only on the effective $T_2^*$ time and occurring incoherent processes alone (up to 600 nW).

\begin{figure}[t!]
\centering
\includegraphics[width=1.15\linewidth]{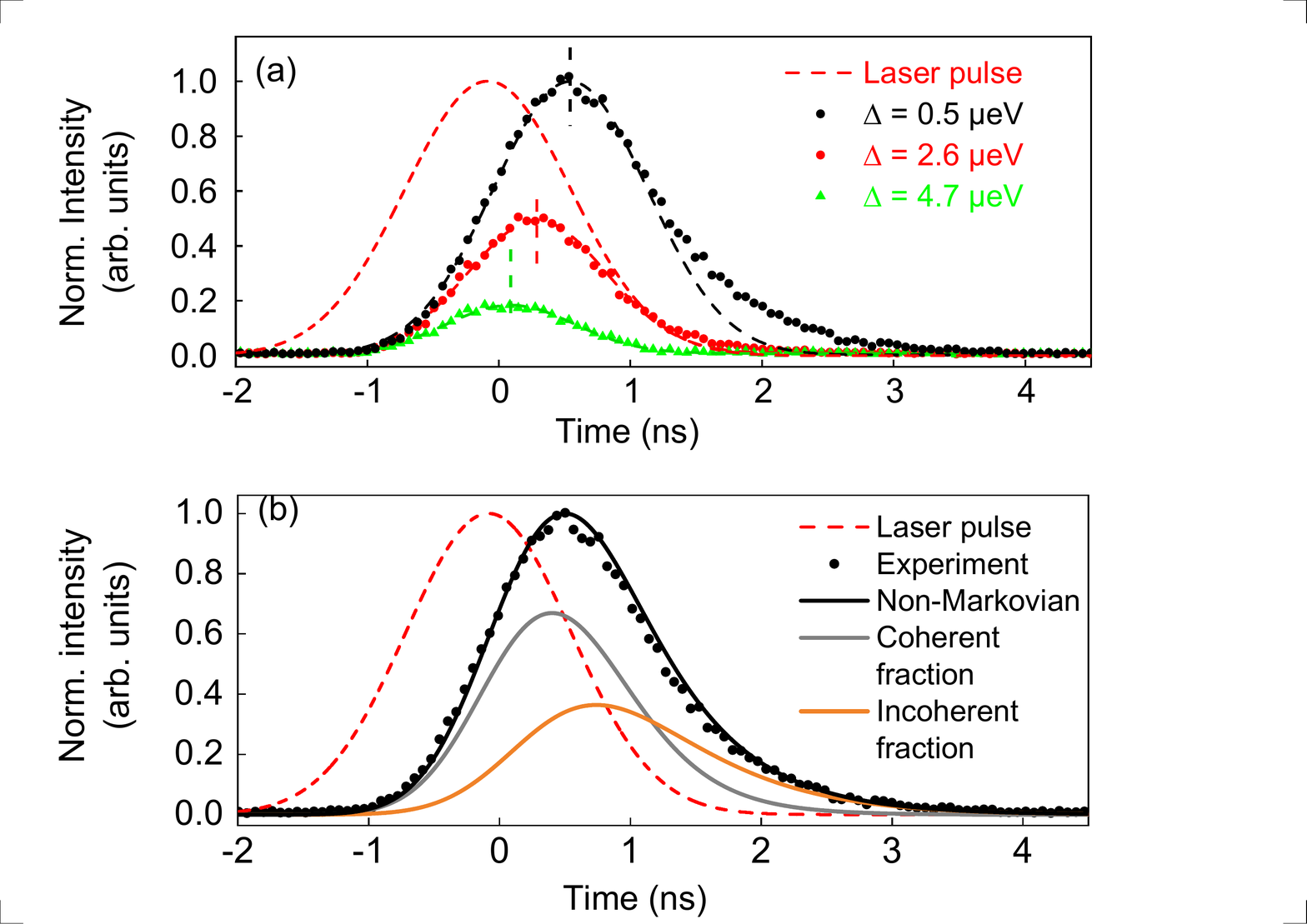}
\caption{\label{graph:2} (a) Scattered pulses for three different laser detunings (Black: 0.5\,$\mu$eV, Red: 2.6\,$\mu$eV, Green: 4.7\,$\mu$eV). (b) Simulation of the pulse detuned by 0.5\,$\mu$eV. The non-Markovian simulation (black line) allows one to identify coherent (grey line) and incoherent contributions (orange line).}
\end{figure}

Staying deep in the Heitler regime, we now focus on details of the time response of the pulse driven QD TLS. Fig.~\ref{graph:2}~(a) shows the scattered pulses for three different detunings. The pulse maxima are determined by fitting Gaussian pulses to the central area of the experimental data. Evidently, the emitted pulses exhibit a deviation from the Gaussian shape on the tail for $t>0$. 

Fitting the emitted pulse shapes in Fig.~\ref{graph:2}(a) via our model considering the microscopic electron-phonon interaction allows us to extract an effective pure dephasing time and gives insight into the semiconductor environment.
In Fig.~\ref{graph:2}~(b), we compare the measured signal for a detuning of $\Delta = 0.5\ \mu$eV (black dots) with the non-Markovian simulation of the population dynamics $\left\langle a^\dagger_c a^{\phantom{\dagger}}_c\right\rangle$ (black line). Best agreement between experiment and theory is obtained 
using $T_1$ = 600 ps and a standard GaAs phonon parameters (given in the SI).
As expected, the coherent fraction 
$|\left\langle a^\dagger_v a^{\phantom{\dagger}}_c\right\rangle|^2$ (grey line) is the dominant source of the signal initially, but for longer times the incoherent
fraction $\left\langle a^\dagger_c a^{\phantom{\dagger}}_c\right\rangle - |\left\langle a^\dagger_v a^{\phantom{\dagger}}_c\right\rangle|^2$
(orange line) takes over due to phonon-assisted fluorescence: $\left\langle a^\dagger_v a^{\phantom{\dagger}}_c\right\rangle \rightarrow \left\langle a^\dagger_v a^{\phantom{\dagger}}_c b^{\dagger}_{\bf 
q} \right\rangle \rightarrow \left\langle a^\dagger_c a^{\phantom{\dagger}}_c b^{\dagger}_{\bf q} \right\rangle$.
Interestingly, this indicates that the semiconductor environment introduces coherent as well as incoherent dynamics into the systems since they open additional excitation and relaxation channels  between the coherence and the excited density \cite{hughes,PhysRevLett.110.113604} 
(see set of equation of motion in SI).
The Wigner delay measurement reveals directly this interesting and important phonon-induced aspect, where the observed non-Gaussian shape in Fig.~\ref{graph:2} is a direct consequence of the electron-phonon interaction in a sense that phonons scatter between the dressed states of the emitter and elongate the emission processes by the suppression of laser-induced coherence. 

\begin{figure}  
\includegraphics[width=\linewidth]{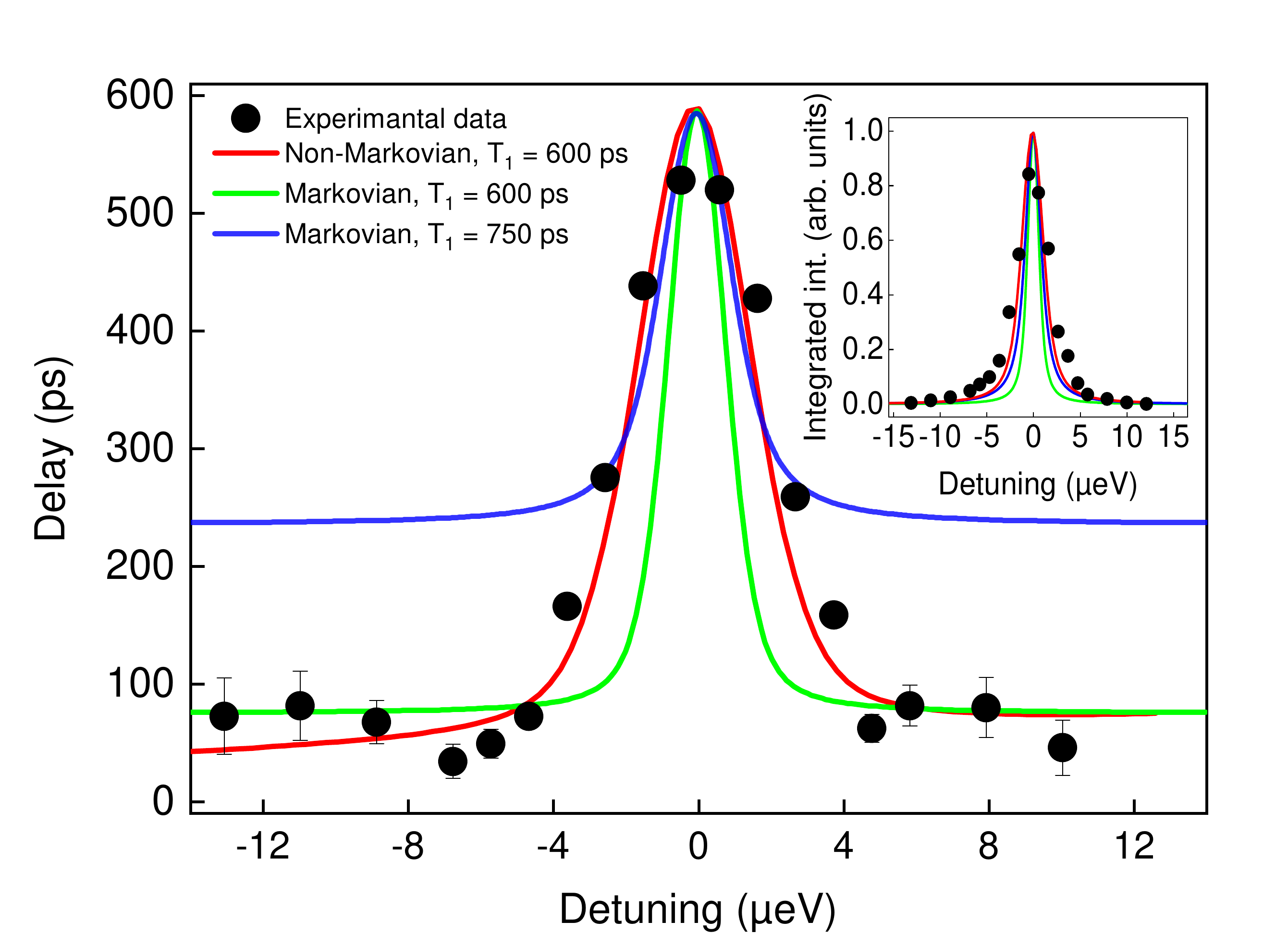}
\caption{Time delay as a function of spectral detuning between the laser and the TLS. The experiment (black bullets) is well described by our model taking non-Markovian dynamics into account (red line), while simple Markovian dynamics (blue and green lines) fails to reproduce the experimental data. Inset: Integrated intensity of the scattered pulses as a function of detuning together with simulations with the same color code as in the main part of the figure.}
 \label{graph:3}
\end{figure}

Another interesting parameter which influences the Wigner time delay is the spectral detuning $\Delta$ between the laser and the TLS. In Fig.~\ref{graph:3}, we plot the experimental time delay (black dots) for different detunings of the laser relative to the exciton transition. The time delay has a maximum of 530\,ps at resonance with direct conversion of laser coherence to emitter coherence and decreases by detuning the laser relative to the exciton transition when indirect conversion occurs.
In atomic systems, the Wigner delay drops to zero for a detuning larger than the radiative linewidth. A solid-state emitter is, however, exposed to electron-phonon interaction which introduces an off-resonance feeding mechanism via emission and absorption of phonons into the dynamics. Such an emitter exhibits therefore remarkably large Wigner delays even for off-resonant driving.
In fact, the observed detuning dependence of the Wigner delay gives access to the coherence properties and allows to evaluate the impact of incoherent processes directly. We propose therefore to take the maximum Wigner delay time for a given setup with a known $T_1$ time as a figure of merit for the effective pure dephasing present in the nanostructure. 

As can be seen in Fig.~\ref{graph:3}, a global Markovian, i.e. frequency independent, pure dephasing constant \cite{carmichael2013statistical,Kreinberg2018} is not sufficient to describe the experimental data. 
In case of a Markovian process, reservoir time correlations are neglected and an information backflow from the reservoir to the system is not considered \cite{breuer2002theory}.  
To reproduce the experimental data for small detunings (blue line) a radiative lifetime of 750 ps and a $T_2^*$ time of 820 ps needs to be assumed, whereas for large detunings (green) a radiative life time of 600 ps and a $T_2^*$ time of 950 ps.
To model both limits with the same experimentally consistent radiative life time of 600 ps a non-Markovian model (red line) must be employed.
The non-Markovian theory interpolates well between both limits and captures also the slightly larger dephasing for blue detuned ($\Delta>0$) excitation scenarios.
This is due to the fact that in the low temperature limit phonon emission is favored over phonon absorption, which renders phonon emission-assisted coherences $\left\langle b^\dg_{\bf q} a^\dg_v a^\ndg_c \right\rangle$ stronger and enhances the dephasing for blue detuned driving, accordingly.
Interestingly, the integrated intensity as a function of detuning presented as inset in Fig.~\ref{graph:3} is also best described by taking non-Markovian dynamics into account (red line).
The discrepancy between Markovian and non-Markovian dephasing in Fig.~\ref{graph:3} clearly indicates that the system's dynamics is influenced by memory effects introduced via a frequency-dependent coupling to the electron-phonon interaction $g^{\bf q}_{vc}$ \cite{moelbjerg2012resonance,kabuss2011microscopic,ulhaq2013detuning,PhysRevB.95.201305,PhysRevLett.118.233602}. 

In conclusion, we have demonstrated for the first time the Wigner time delay for a solid state two-level system. The 	Wigner time delay is caused by the phase lag induced by light scattering on a semiconductor QD in a resonance fluorescence experiment at cryogenic temperature. We observe detuning-dependent time delays of up to 530\,ps at resonance between the exciting laser and the two-level system in good agreement with a theoretical description based on the optical Bloch equations. Here, the delay effect is observable despite the inevitable pure dephasing processes in the solid state system which limits the coherent response of the system. Our results show that using high-quality sample fabrication and advanced spectroscopic tools, effects so far limited to the realm of atomic quantum optics can nowadays also be explored using optimized solid state systems. It would be interesting the study the Wigner time delay also in QD-microcavity systems operating in the regime of cavity quantum electrodynamics. Experiments in this regime could provide an opportunity to tailor the spontaneous emission lifetime, and thus the dynamics of resonant light scattering, via the Purcell effect~\cite{Bayer2001}. This may allow one to observe larger Wigner time delays close to a few tens of ns as reported for a single-atom system~\cite{Bourgain2013}. The observed Wigner time delay may be used as a method for temporal fine-tuning in future photonic quantum technology systems such as quantum repeater networks which require a very precise spectral and temporal matching of resonantly excited single-photon emitters for entanglement distribution via Bell-state measurements.

\section*{Acknowledgements}
The research leading to these results has received funding from the 
European Research Council (ERC) under the European Union's Seventh Framework ERC 
Grant Agreement No. 615613 and from the German Research Foundation Project (DFG) No. 
RE2974/5-1. A.C. and J.S. gratefully acknowledge the support by the DFG through project B1 of the SFB 
910. The W{\"u}rzburg group is grateful for support by the State of Bavaria. 

\bibliographystyle{apsrev4-1}

%

\end{document}